\documentclass[runningheads]{llncs}   

\usepackage[T1]{fontenc}

\usepackage{tabu}                      
\usepackage{booktabs}                  
\usepackage{lipsum}                    
\usepackage{mwe}                       
\usepackage{graphicx}
\usepackage{hyperref}
\graphicspath{{figures/}{./}} 

\usepackage{amsmath} 
\usepackage{colortbl}
\usepackage{orcidlink}
\newcommand{\myrowcolour}{\rowcolor[gray]{0.95}}

\usepackage{mathptmx}                  

\begin{document}

\title{Geometry-Aware Global Feature Aggregation for Real-Time Indirect Illumination}


%


\author{Meng Gai\inst{1,2} \and
Guoping Wang\inst{1,2} \and
Sheng Li*\inst{1,2}\thanks{Corresponding author}}

\authorrunning{Gai M. et al.}
%
\institute{School of Computer Science, Peking University, Beijing, China \and
National Key Laboratory of Intelligent Parallel Technology, Beijing, China\\
\email{\{gaimeng,wgp,lisheng\}@pku.edu.cn}}


\maketitle

\begin{abstract}
    Real-time rendering with global illumination is crucial to afford the user realistic experience in virtual environments. We present a learning-based estimator to predict diffuse indirect illumination in screen space, which is then combined with direct illumination to synthesize globally-illuminated high dynamic range (HDR) results. Our approach tackles the challenges of capturing long-range/long-distance indirect illumination when employing neural networks and is generalized to handle complex lighting and scenarios.
    From the neural network thinking of the solver to the rendering equation, we present a novel network architecture to predict indirect illumination. Our network is equipped with a modified attention mechanism that aggregates global information guided by spatial geometry features, as well as a monochromatic design that encodes each color channel individually.  
    We conducted extensive evaluations, and the experimental results demonstrate our superiority over previous learning-based techniques. Our approach excels at handling complex lighting, such as varying-colored lighting and environment lighting. It can successfully capture distant indirect illumination and simulate the interreflections between textured surfaces well (i.e., color bleeding effects); it can also effectively handle new scenes that are not present in the training dataset.  

\end{abstract}

\begin{figure*}
 \includegraphics[width=\linewidth]{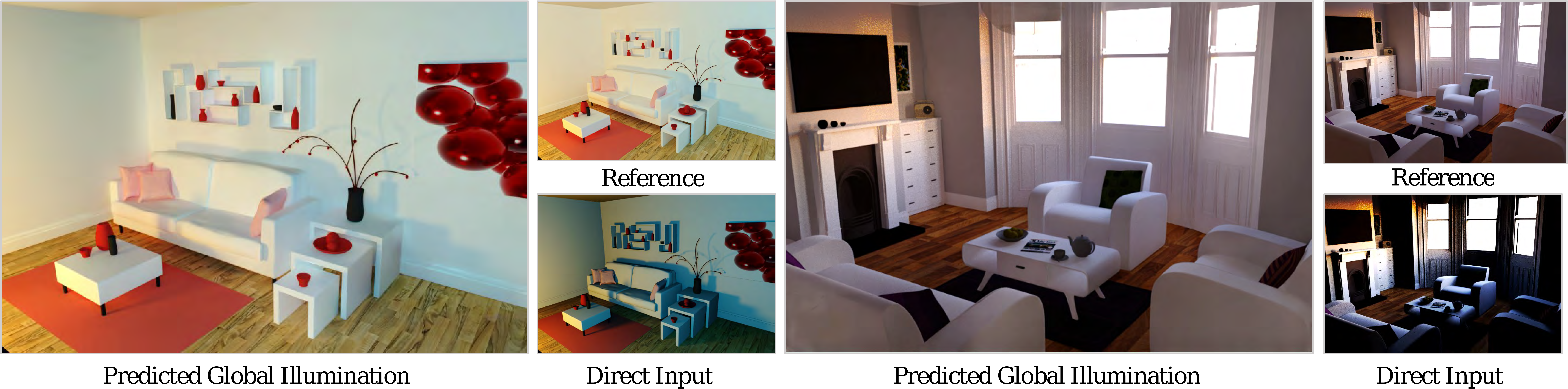}
 \centering
  \caption{ 
  Our model efficiently synthesizes HDR results with global illumination predicted in around 12 milliseconds. We highlight that our model is well generalized to handle complex dynamic lighting in scenes, such as varying-colored lighting (left) and environment lighting (right), which are new scenes that have not been present in the training dataset. }
  \label{fig:teaser}
\end{figure*}





\section{Introduction}

\label{sec:intro}
Visual realism with global illumination is a key demand of immersive virtual reality (VR) applications to intrigue users' immersion arousal and respond realistically to virtual environments and
events \cite{yu2012visual}. However, realistic rendering of such globally-illuminated scenarios in real time remains a big challenge, especially in handling complex lighting situations like dynamic changes of illumination and environment lighting \cite{mortensen2008real}.
While photo-realistic rendering techniques (e.g., Monte Carlo path tracing) can synthesize images with global illumination, they are often extremely time-consuming and may take hours to render. Meanwhile, various works attempt to use approximations along with classical rasterization-based rendering, thus approximating global illumination effects at interactive or real-time performance. They generally operate in either image or 3D space and adopt different precomputation techniques to reduce runtime cost~\cite{sloan2002prt,keller1997instant}. 
Recently, learning-based techniques for photo-realistic rendering were proposed, some of which combine data-driven methods with classic rendering pipeline, resulting in improved image quality or rendering speedup~\cite{dai2020pbr,vicini2019learned}. These techniques, namely neural rendering~\cite{tewari2020state}, usually provide stronger supervision to generative models by applying the prior knowledge in classical rendering techniques to neural networks.

Predicting indirect illumination is, nonetheless, a challenging task. Prior works generally regard this as a special case of image-to-image translation, and use convolutional neural networks (CNN) for indirect/global illumination prediction. The auxiliary geometry features are provided as inputs to predict global illumination, further conditioned on input the direct lighting~\cite{nalbach2017deep,xin2021lightweight} or emission maps~\cite{dai2020pbr}. However, several major challenges remain: 
\begin{itemize}
    \item Poor generalization. Various lighting conditions, geometries, and surface textures jointly result in diversified scenes, inflicting high demands on both training data and model capacity. However, synthetic datasets have limited coverage; and the need for real-time performance restricts model capacity/complexity.
    \item Difficult to model long-distance indirect illumination due to narrow-range dependencies. Prior works generally use stacks of convolutional operations within their model, where the processing of inputs is limited to a local neighborhood.
    \item Inaccurate color when predicting indirect illumination. We find that the generative models easily fall into predicting the average of the distribution within the training data, resulting in inconsistency between direct and indirect illumination.
\end{itemize}

To address these issues, we adopt a learning-based approach to address the issue of global illumination, following the principles of the rendering equation~\cite{kaj86re}. Our approach includes a learnable module that aggregates global features while utilizing geometry supervision. This module incorporates a modified multi-head attention mechanism~\cite{vaswani2017attention} to calculate the distribution based on geometry features, which guides the global aggregation of spatial features. This non-local process can be analogized to an integral over screen space, where the geometry-aware attention works as a learnable counterpart of the geometry term. Our proposed module overcomes the network bottleneck of the deep stack of convolutional layers, requiring less computation and fewer parameters while better capturing the long-range dependencies. Moreover, it can be applied to other tasks where geometry supervision is available and long-range dependency modeling is preferred.

Second, we present a monochromatic model design, where each RGB channel is independently processed using a shading generator conditioned on shared features from a geometry encoder. 
The redundant inter-channel interactions (introduced by the interconnections of convolutional filters) can be eliminated, and this relieves the burden of training data and network capacity, and enforces the network to learn the accurate color of indirect light.
Furthermore, this makes our model immune to color variations, therefore allowing better generalization to novel and complex scenes and dynamic lighting, which are not present in the training data. 
Since HDR images can provide a much wider range of brightness and contrast for the high sensitivity of the human visual system to enhance the user's perception of immersion and presence in the immersive virtual environment \cite{GIHDR2008,HDRVR2022}, our model is trained to synthesize HDR global illumination.
 
Finally, we created a novel synthetic HDR dataset of indoor scenes for network training based on the public indoor layout dataset 3D-Front~\cite{fu20203dfront}. Our dataset comes with well-annotated lighting, texture, and geometry features; the surfaces are also assigned with physically based materials, providing sufficient diversity for geometry features and visual appearances. We design an adversarial training framework to directly optimize our network on HDR data, where the discriminator and the perceptual loss are jointly used for network optimization. 

Our model can produce plausible indirect illumination at real-time performance, handling global effects, accurately predicting indirect shading with good generalization, as shown in Fig.~\ref{fig:teaser}.
To summarize, our main contributions include: 
\begin{itemize}
    \item We propose a neural network for indirect illumination prediction that leverages a global feature aggregation module for geometry features to model long-range dependencies, thereby synthesizing realistic HDR images with high performance.
    
    \item We present a monochromatic design featuring the independent processing of each color channel, which benefits our model in terms of efficiency, compactness, and explainability. 

    \item We create a photo-realistic synthetic HDR dataset with pixel-wise annotated features, including factorized lighting, material, and geometry. Our dataset can offer significant potential as a resource for learning-based rendering algorithms to facilitate diverse benchmarks of real-time global illumination, supporting high presence in immersive virtual reality applications.
\end{itemize}

\section{Related Work}
\label{sec:related}
\subsection{Real-Time Global Illumination} 
Numerous works seek to acquire global illumination effects at a low time budget. Classical methods adopt different approximations to reduce performance costs. Some of them operate in 2D screen space, such as screen space ambient occlusion (SSAO)~\cite{bavoil2008image} and screen space directional occlusion (SSDO)~\cite{ritschel2009ssdo}, using rasterized direct lighting and g-buffer features to approximate inter-reflections between surfaces. While some others operate in 3D space, usually achieving more realism with higher computational costs and storage, such as voxel cone tracing (VXGI)~\cite{crassin2011vxgi} and light propagation volumes (LPV)~\cite{kaplanyan2009lpv}. Lastly, precomputation-based methods like precomputed radiance transfer (PRT)~\cite{sloan2002prt} reduce the runtime cost by computing pre-integrations on the environment and BRDFs. Recently, with the advances in deep learning, data-driven methods have also been adopted to improve rendering quality efficiently. For example, training CNNs to denoise low sample-count images~\cite{chaitanya2017mc}, or predicting the proper sample count for real-time raytracing~\cite{kuznetsov2018deep}.

\subsection{Image-to-Image Translation}
The task of image-to-image translation converts images from one domain to another, e.g., image style translation, colorization, etc. The architecture of convolutional neural networks (CNN) has been the most commonly used technique for image-to-image tasks, such as U-Net~\cite{ronneberger2015u} and ResNet~\cite{he2016deep}. After the generative adversarial network (GAN) proposed by Goodfellow et al.~\cite{goodfellow2014generative}, it has been further extended to conditional generative models and introduced to image-to-image tasks as well~\cite{isola2017image}. Recently, there has been another trend of using pre-trained models (e.g., VGG-19~\cite{simonyan2014vgg}) to capture structurally correlated high-level image features for network optimization, namely the perceptual loss~\cite{johnson2016perceptual}. We show that a model's output quality and robustness could be enhanced by jointly introducing adversarial loss and perceptual loss into network optimization. 

\subsection{Network Conditioning}
Image-to-image translation could also be regarded as conditional synthesis, where networks learn to generate images conditioned on input geometry or semantic labels. Most of the early works simply provide conditioning data as the input of the network~\cite{isola2017image}. In contrast, many recent works focus on finding better ways to condition the image generation process, e.g., using a depth map to explicitly guide the convolution~\cite{wang2018depth}, using learned transformations to modulate convolutional features~\cite{park2019semantic}. While these works have achieved great success, they are mainly used along with convolution layers, and thus remain local conditioning methods. 

Inspired by the great success of attention mechanisms in modeling global dependencies~\cite{vaswani2017attention,zhang2019self} and motivated by our task formulation, we propose a feature aggregation method that models long-range dependencies guided by geometry features, where the aggregation weights are obtained from a modified multi-head attention mechanism. Different from the prior work on integrating attention to CNN~\cite{zhang2019self}, we 1) explicitly use geometry features to guide the aggregation of global features; then 2) extend it into multi-head attention~\cite{vaswani2017attention}, and find it sufficient to equip only bottleneck layers with attention. Both bring a significant reduction of computational overhead, with no observable performance degradation in our experiments.  

\subsection{Global illumination prediction} 
Several recent works use neural networks to predict global illumination given auxiliary shading and geometry features. Nalbach et al.~\cite{nalbach2017deep} first introduced CNN into global illumination prediction. They use a U-shaped CNN, taking pixel-wise features as inputs and predicting globally illuminated images. Bi et al.~\cite{bi2019deep} train GANs to predict global illumination, and Dai et al.~\cite{dai2020pbr} use the perceptual loss for network optimization. Then Xin et al.~\cite{xin2021lightweight} used a lightweight CNN and predicted the indirect component similar to the task setup in our work. However, Nalbach et al.~\cite{nalbach2017deep} used global illumination computed in screen space as their ground truth, and Dai et al. excluded textures in their ground truth data generation. Most prior works used LDR data for training or took no special care in training in linear HDR space. Moreover, they generally use stacked convolutional operations for shading learning, limiting the receptive fields to local neighborhoods, and thus it is difficult to predict global effects like indirect illumination.  

Our approach focuses on capturing more accurate long-range indirect illumination while better handling the inter-reflections between differently textured surfaces (i.e., color-bleeding effects).

\begin{figure*}[!ht]
  \centering
  \includegraphics[width=\linewidth]{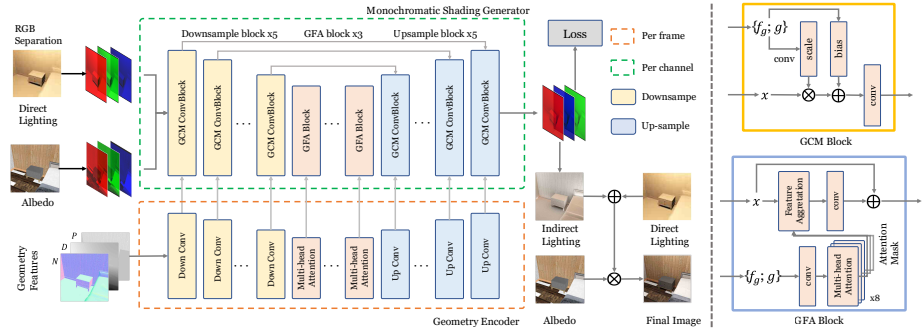}
  \caption{\label{fig:network}%
       Overview of our model architecture. GCM: geometry-aware conditional modulation; GFA: geometry-aware feature aggregation. The geometry encoder takes in the auxiliary geometry information of each frame as input and provides geometry conditional features (GCM) and attention weights (GFA) to the generator. The monochromatic shading generator then independently predicts each color channel of the indirect illumination, given the lighting and reflectance of the corresponding channel as input.}
\end{figure*}

\section{Our Method}
\label{sec:method}
We first outline the neural network thinking of the solver for rendering equations. Then, we present our network architecture, including the feature aggregation module and the monochromatic design. Finally, we discuss the technical details, practical considerations about performance and optimization, and the adversarial training framework.

\subsection{Task Formulation}
We aim to predict indirect lighting given direct lighting and auxiliary geometry features, including surface normal, depth, reflectance, etc. All these inputs can be efficiently computed using a rasterization-based rendering pipeline in real time. For simplicity, we assume all Lambertian surfaces within the scene, i.e., the BRDF term $f(\mathrm{p},\omega_i, \omega_o)$ is a constant for arbitrary $\omega_i$ and $\omega_o$. We first briefly review the concepts of rendering equation~\cite{kaj86re} and physically based rendering; by splitting the incident radiance $L_i$ into direct and indirect terms, we derive an alternative form of the surface rendering equation:
\begin{equation}
\begin{aligned}
   L_o(\mathrm{p},\, \omega_o) & = \int_{A}\left(L_i^{\text{d}}(\mathrm{p},\, \omega_i)+L_i^{\text{ind}}(\mathrm{p},\, \omega_i)\right)f_r(\mathrm{p})G(\mathrm{p},\, \mathrm{p^\prime}) \,\mathrm{d}A(\mathrm{p^\prime}) \\
   &= L_o^{\text{d}}(\mathrm{p}, \,\omega_o) + L_o^{\text{ind}}(\mathrm{p},\, \omega_o) \ ,
\label{equation:re}
\end{aligned}
\end{equation}

\noindent
where ${L}_\mathrm{o}^\mathrm{d}$ and ${L}_\mathrm{o}^\mathrm{ind}$ are the direct and indirect components of the scattering radiance, respectively. $G(\mathrm{p},\, \mathrm{p^\prime}) = \frac{ V(\mathrm{p^\prime} \leftrightarrow \mathrm{p}) {\cos\theta \cos\theta^\prime} } {\parallel\mathrm{p} - \mathrm{p^\prime} \parallel}$ is the geometry term where $V(\mathrm{p^\prime} \leftrightarrow \mathrm{p})$ indicates the visibility between two vertices.
We can obtain the direct lighting term ${L}_\mathrm{o}^\mathrm{d}$ by integrating over all light sources or by real-time alternatives~\cite{sloan2002prt,fernando2005pcss,heitz2016real}. We then feed ${L}_{\mathrm{o}}^{\mathrm{d}}$ into the network and predict the indirect component ${L}_{\mathrm{o}}^{\mathrm{ind}}$.

To apply fine-grained supervision to the network, we first decompose the direct and indirect components of the final illumination, letting the network focus on predicting the indirect component instead of global illumination. We then use reflectance factorization to demodulate shading and diffuse reflectance, thus effectively tasking the network with learning the demodulated shading (instead of scattered radiance), which has a relatively low frequency after multiple bounces within the scene. The final rendering is obtained by multiplying the predicted shading with the reflectance. Moreover, we propose a feature aggregation module for better modeling global dependencies, which is intuitively similar to an integral over screen space. We employ the modified attention mechanism~\cite{vaswani2017attention} to impose the effects of distance and visibility on network models, which intuitively works like a learnable counterpart of the geometry term, as is discussed later.

\noindent
\textbf{Summarization} To predict indirect illumination, the geometry information (surface normals $\mathbf{N}$, depth $\mathbf{D}$, diffuse reflectance $\mathbf{R}$, bold type indicates image data) and direct lighting $\mathbf{L}_{\mathrm{d}}$ are provided as conditioning features. The network is expected to predict the indirect shading component $\mathbf{S}_{\mathrm{ind}}$, from which we can obtain the full indirect illumination $\mathbf{L}_{\mathrm{ind}}$ by multiplying the reflectance, i.e., $\mathbf{L}=\mathbf{R} \cdot \mathbf{S}$. The final global illumination $\mathbf{L}$ is acquired by adding direct illumination $\mathbf{L}_{\mathrm{d}}$ (provided as inputs) to the predicted indirect illumination, i.e., $\mathbf{L} = \mathbf{L}_{\mathrm{d}} + \mathbf{L}_{\mathrm{ind}} $. Note that all the above computations are pixel-wise operations in linear space; thus, all the components should be represented with linear values and stored as HDR images. 

\subsection{Model Design}
\textbf{Overview} 
Our model contains a shading generation branch and a geometry encoding branch, both using a U-shaped network with skip-connections~\cite{ronneberger2015u}. 
We integrate 1) our feature aggregation module into the shading generator to model long-range dependencies; 2) the monochromatic design for a more consistent deep rendering process.
We also utilize conditional modulation techniques~\cite{park2019semantic} for better geometry-aware supervision. The overview of our network architecture is shown in Fig. \ref{fig:network}.

\noindent
\textbf{Geometry Encoder} 
We use an additional geometry encoder that encodes geometry properties (i.e., positions $\mathbf{P}$, normals $\mathbf{N}$, and depth $\mathbf{D}$), leaving the other inputs to the shading encoder. The encoded geometry features are provided to condition the synthesis process of the shading generator. The separation of the encoding process splits the task into spatial structure modeling and light transport simulation, thus alleviating the requirements on training data and network complexity.
Instead of simply concatenating the geometry features to the inputs of the generator layers, we use a more effective method, namely geometry-aware conditional modulation (GCM) similar to~\cite{park2019semantic}. Specifically, the conditional geometry features are transformed into scaling and biasing factors through learnable convolution layers as:
\begin{equation}
\mathbf{GCM}(x) = \left(\gamma(\hat{f_g}) \otimes {x}\right) \oplus \beta(\hat{f_g}),
\end{equation}
where $\oplus$ and $\otimes$ denote element-wise addition and multiplication, respectively. $\hat{f_g}=\{ f_g ;\, g \}$ denotes the layer-wise encoded geometry features concatenated with original geometry inputs (i.e., $\mathbf{N}, \mathbf{D}$, etc.), while $\gamma$ and $\beta$ are learnable transformations implemented with convolution layers. Experimentally, this is more effective and computationally saving when jointly used with our monochromatic network, as discussed below.

\noindent
\textbf{Geometry-Aware Feature Aggregation}
Fully convolutional networks (FCN) will result in a limited-sized receptive field~\cite{xin2021lightweight,nalbach2017deep,dai2020pbr}, i.e., the output pixel value is conditioned only on a local region. While deep stacks of convolutional operations can enlarge the receptive field, they also bring prohibitive computational costs, making network optimization difficult. In this work, we expect the network to generate indirect illumination from features of all locations within the image space. A learnable module is needed that can efficiently aggregate global features guided by geometry.

\begin{figure}[t]
    \begin{center}
    \includegraphics[width=0.65\linewidth]{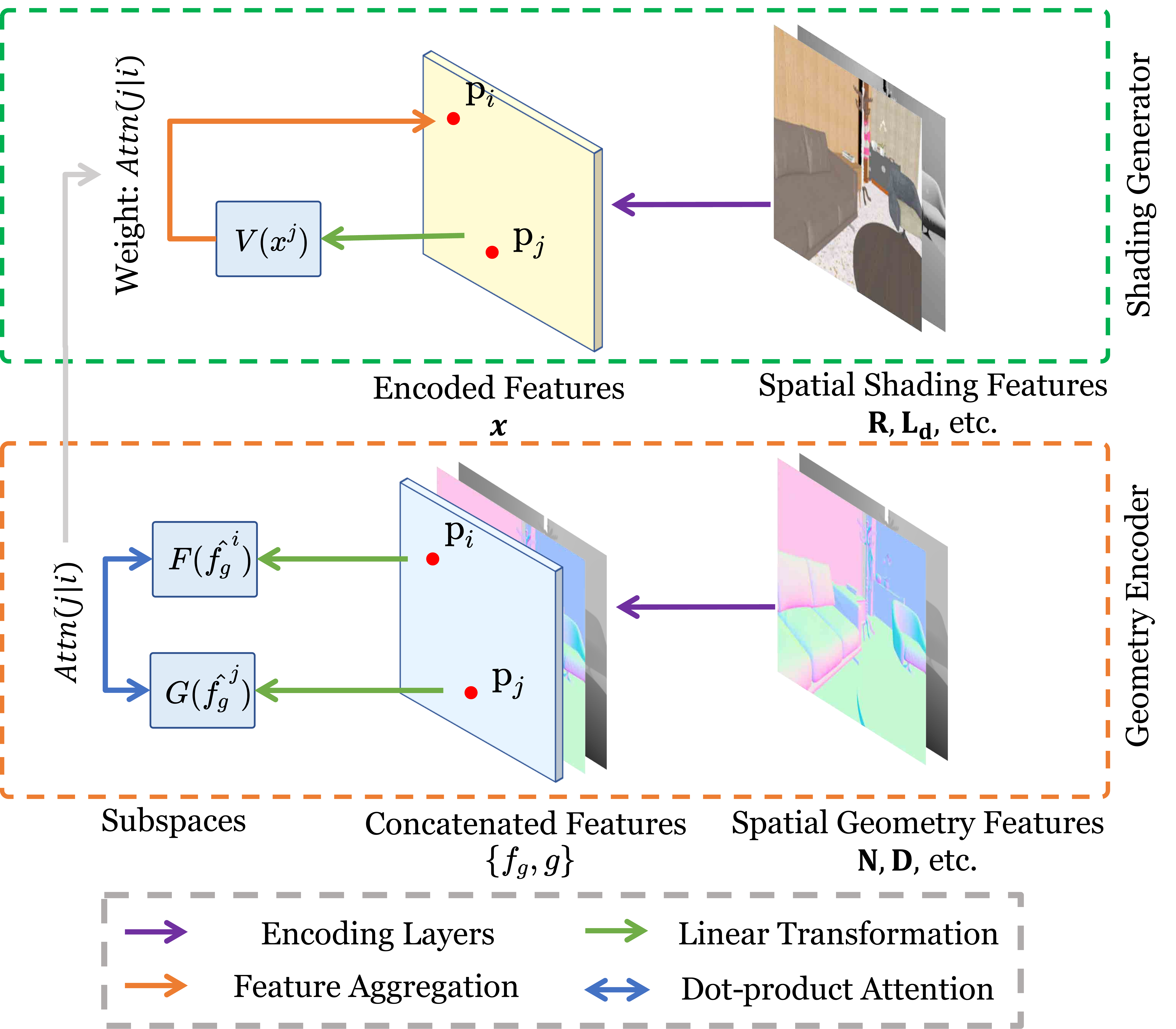}  
    \caption{\label{fig:gfa}
         Geometry-aware feature aggregation operation. $\mathrm{p}_i$ and $\mathrm{p}_j$ could be any two 2D locations of the input features, thus effectively modeling global dependencies within screen space.}
    \end{center}
\end{figure}

We use a modified multi-head attention mechanism~\cite{vaswani2017attention} to acquire the weights for the feature aggregation module based on geometry features (i.e., positions, depth, and surface normals). 
For each spatial location, the module first calculates the attention weights based on the encoded geometry features $f_g$, then obtains the response of that position as a weighted combination of the features from all locations. The attention weights $\mathrm{Attn}(j\,|\,i)$ for location $i$ are computed as:
\begin{equation}
    \mathrm{Attn}(j\,|\,i) = \mathrm{Softmax}_j\left(\mathrm{Dot}\left(Q(\hat{f_g}^i), K(\hat{f_g}^j))\right)\right),
\end{equation}
where $i$ and $j$ are indices to the features within image space, $Q$ and $K$ are linear transformations (implemented as $1\times 1$ convolutions). Specifically, $F$ and $G$ correspond to these two transformations in the geometry encoder, representing the Query ($Q$) and Key ($K$) projections applied to geometry features, as shown in Fig.~\ref{fig:gfa}. The encoded features $\hat{f_g}$ are again concatenated with original geometry inputs (down-sampled to the current size) to preserve them from being washed away. The attention distribution of $i$ is normalized for all spatial location $j$ using $\mathrm{softmax}$, and is then used to get the aggregated global feature $\hat{x}^i$ for location $i$ as:
\begin{equation}
    \hat{x}^i = x^i + \Sigma^N_{j=1}\left(V(x^j)\,\mathrm{Attn}(j\,|\,i) \right),
\end{equation}
where $x^i$ is the residual connection applied to outputs, and $V$ is an extra linear transformation applied to the original features while preserving their original dimensionality. Intuitively, this process is similar to the process of integrating over screen space, where the attention mechanism calculates some sort of mutual contribution between two points for shading based on their geometry features (just like a learnable counterpart of the geometry term, Eq.~\ref{equation:re}). 
By attending to the projected subspace of geometry features, it learns to determine whether the feature $x_j$ of location $j$ is useful and thus should be aggregated into the current location $i$ (see Fig.~\ref{fig:gfa}).

In addition, we find it beneficial to use multiple learnable linear transformations within attention computation, i.e., obtain multiple attention distributions $\mathrm{Attn}_k$ using different sets of linear transformations $Q_k$, $K_k$, and $V_k$, where $k=1...h$ is the index of attention head and $h=8$ in our experiments. The aggregated features $\hat{x_k}$ are then concatenated to produce the final output $\hat {x}$, resulting in the multi-head version~\cite{vaswani2017attention} of the aggregation module. This allows the network to simultaneously attend to different subspaces of the spatial features, resulting in a better capacity for modeling long-range dependencies.

\noindent
\textbf{Monochromatic Shading}
The key insight of the monochromatic design is from the difference between the convolutional neural networks (CNN) and a typical rendering pipeline. The CNN generally encodes an RGB image in one pass, where the features in different color channels could interact via the inter-channel connections in convolutional filters. However, this is inconsistent with a typical rendering pipeline, where the rendering process of each color channel (or the light transport of a specified wavelength) is handled independently. 

Specifically, we interpret the rendering process as a mapping $F: \mathcal{X} \to  \mathcal{Y}$, where $\mathcal{X}$ is the high-dimensional inputs encoding the scene geometry, camera, materials, etc., and $\mathcal{Y}$ is the rendered 2D image, thus, rendering can be viewed as evaluating the function $Y=F(X)$. Since the rendering process is wavelength-independent (thus is channel-independent if the RGB color model is used), this function can be further written in a channel-independent way, i.e., $Y^c = F^c(X^c)$, where $c\in \{\mathrm{r},\mathrm{g},\mathrm{b}\}$ indicates the specific channel. 
However, in the general learning-based methods, the mapping between inputs and outputs is written as $\hat Y = G(X;\theta)$, where $\theta$ is the model parameter, and $\hat Y$ is the predicted indirect illumination. Note that each color channel of the RGB inputs (i.e., reflectance and lighting) contributes to all three channels of $\hat Y$ via the interconnections within the convolutional filters, which is redundant and inconsistent with the typical rendering process.

Based on this observation, our monochromatic shading generator processes each color channel independently, i.e., for each input with RGB channels, the shading network takes in each channel in parallel and predicts the shading of that channel. Note that the same shading network is shared between different color channels. We could thus interpret the prediction process as: 
\begin{equation}
    \mathbf{\hat L}_{\mathrm{ind}}^c = \mathbf{Generator}(\mathbf{L}_{\mathrm{d}}^c, \mathbf{R}^c, f_g; \,\theta) \ ,
\end{equation}
where $c$ indicates the current color channel and $\mathbf{f}_g$ is the encoded features from the geometry encoder. 
This channel-independent process is consistent with the typical rendering process, freeing our network from the inequalities in the color distribution in training data. Moreover, the RGB feature of each pixel is not considered helpful in our task, where we seek to simulate light transportation only, which is channel-wise independent. We will discuss its superiority in Sec. \ref{sec:results}.

\subsection{Optimizations}
\noindent
\textbf{Minimizing Computational Overhead}
The conditional feature modulation (CFM) modules are carefully designed to enhance the data reusing/sharing between channels, thus minimizing the extra computational overhead introduced by the monochromatic design. Specifically, the conditional features ($\gamma(\hat{f_g})$ and $\beta(\hat{f_g})$ by GCM Block), as well as the attention distribution ($\mathrm{Attn}(j\,|\,i)$ by GFA Block) are both only dependent on the geometry features (see Fig. ~\ref{fig:network}), thus are shared and reused by different color channels. While a per-channel process is still needed(instead of directly reconstructing RGB outputs), we later justify this potential loss of efficiency by showing that a smaller monochromatic model can still perform better with similar computational complexity and fewer parameters than a normal model. 

\noindent
\textbf{Optimizing on HDR Images} 
Most image generation tasks use tone-mapped low dynamic range (LDR) images for training and directly render tone-mapped images in LDR space. We note that multiple benefits could be obtained by optimizing the model directly using linear HDR data, including 1) preserving details in both dark bright areas of high contrast images; 2) preserving the zero-mean distribution of noise (introduced by unbiased Monte Carlo sampling), which may be beneficial for the learning process; and 3) allowing for more flexible post-processing applications on outputs, such as re-exposure or refocus. Moreover, if we use tone-mapped LDR images for both input and output, the physical correctness (i.e., the linearity between direct and indirect components) would be lost in our training data. 

To better parameterize the linear radiance values, we use an exponential activation for the generator output rather than a sigmoid function commonly used for reconstructing LDR images.
Furthermore, the linearity between direct illumination and indirect illumination could be utilized with HDR training. Specifically, for the direct $\mathbf{L}_{\mathrm{d}}$ and indirect $\mathbf{L}_{\mathrm{ind}}$ components within a rendered frame, if we perturb the direct illumination with a random factor, i.e., $ \mathbf{\hat{L}}_{\mathrm{d}} = \alpha \cdot \mathbf{L}_{\mathrm{d}}$ (equivalent to re-exposure or scaling all light sources within the scene), we will have its corresponding indirect illumination as $\mathbf{\hat{L}}_{\mathrm{ind}} = \alpha \cdot \mathbf{L}_{\mathrm{ind}}$. We exploit this fact by adding random perturbations to the exposure of each frame while training, which improves the model's robustness to illumination variations. This could also be regarded as an effective data augmentation technique.

\subsection{Adversarial Training and Loss Function}

\begin{figure}[tbp]
    \centering
    \includegraphics[width=0.7\linewidth]{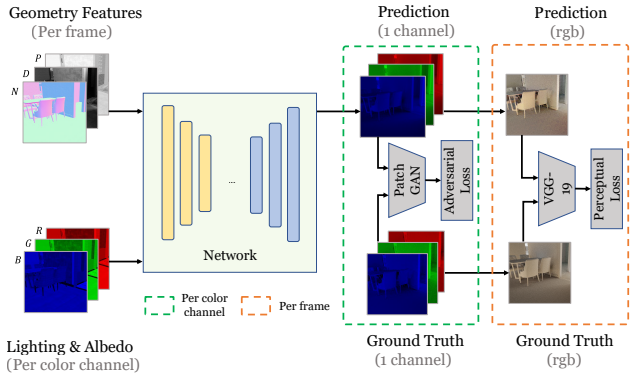}
    \caption{Our adversarial training framework. We use a PatchGAN as the discriminator for adversarial training, and the perceptual loss is acquired using a pre-trained VGG-19. The computation of adversarial loss is in a per-color-channel manner, whereas the perceptual loss is computed on the final RGB image.}
    \label{fig:loss}
\end{figure}

We present an adversarial training framework using a weighted combination of losses for network optimization, including a content loss $ \mathcal{L}_\text{c}$, a perceptual loss $ \mathcal{L}_\text{p}$, and an adversarial loss $ \mathcal{L}_\text{a}$. As illustrated in Fig. \ref{fig:loss}, our final loss function $\mathcal{L}$ can be written as: 
\begin{equation}
\label{equation:loss}
    \mathcal{L} = \omega_{\mathrm{a}} \mathcal{L}_\mathrm{a} + 
    \omega_{\mathrm{c}} \mathcal{L}_\mathrm{c} + 
    \omega_{\mathrm{p}} \mathcal{L}_\mathrm{p} \ ,
\end{equation}
where $\omega_{\mathrm{c}}$, $\omega_{\mathrm{p}}$, and $\omega_{\mathrm{a}}$ are the hyperparameters controlling the weight of each loss term,  we use empirical values where $\omega_{\mathrm{c}}=0.7$, $\omega_{\mathrm{p}}=0.28$, $\omega_{\mathrm{a}}=0.02$ in our experiments.  We use smooth L1 loss~\cite{ren2015faster} as our content loss. While the L1 loss could not capture the inter-pixel correlations or the structural information of the image, we thus additionally introduced a PatchGAN~\cite{isola2017image} as the discriminator to obtain the adversarial loss term, which helps recover high-frequency local details. Moreover, it is an ill-posed problem to predict indirect illumination given direct lighting in screen space only (instead of in 3D solutions). While classic pixel-wise loss (e.g., L1, L2) generally tends to predict the average value over all possible solutions~\cite{ledig2017photo}, we hope the adversarial training can guide the generator to approximate one particular solution, thus being more consistent with human perception.

Furthermore, to bridge the gap between the generative model and human perception, we introduce the perceptual loss~\cite{johnson2016perceptual}, which captures the structurally correlated high-level features and thus is robust to noise and local artifacts. Specifically, we compute the perceptual loss $\mathcal{L}_{\text{p}}$ based on the $\mathrm{L}1$ distance of different convolutional layers within a pre-trained VGG-19~\cite{simonyan2014vgg}:
\begin{equation}
    \label{equation:vgg}
    \mathcal{L}_{\mathrm{p}}(X, Y, \theta) = \sum^N_l \lambda_i \left\Vert \Phi_l(Y) - \Phi_l(\mathbf{G}(X, \theta)) \right\|_1 \ ,
\end{equation}
where ${\Phi_l}$ is the $l\text{-}\mathrm{th}$ layer of VGG-19 and $\lambda_i$ is the weight of each layer. Note that while the content loss and the adversarial loss are computed and averaged for each color channel, the perceptual loss, however, is applied to the composited RGB image (see Fig. \ref{fig:loss}), making the final output coherent with human perception and maintaining the visual consistency across independently processed channels. 

\section{Synthetic Dataset}
\label{sec:dataset}
\begin{figure}[tbp]
    \centering
    \includegraphics[width=0.7\linewidth]{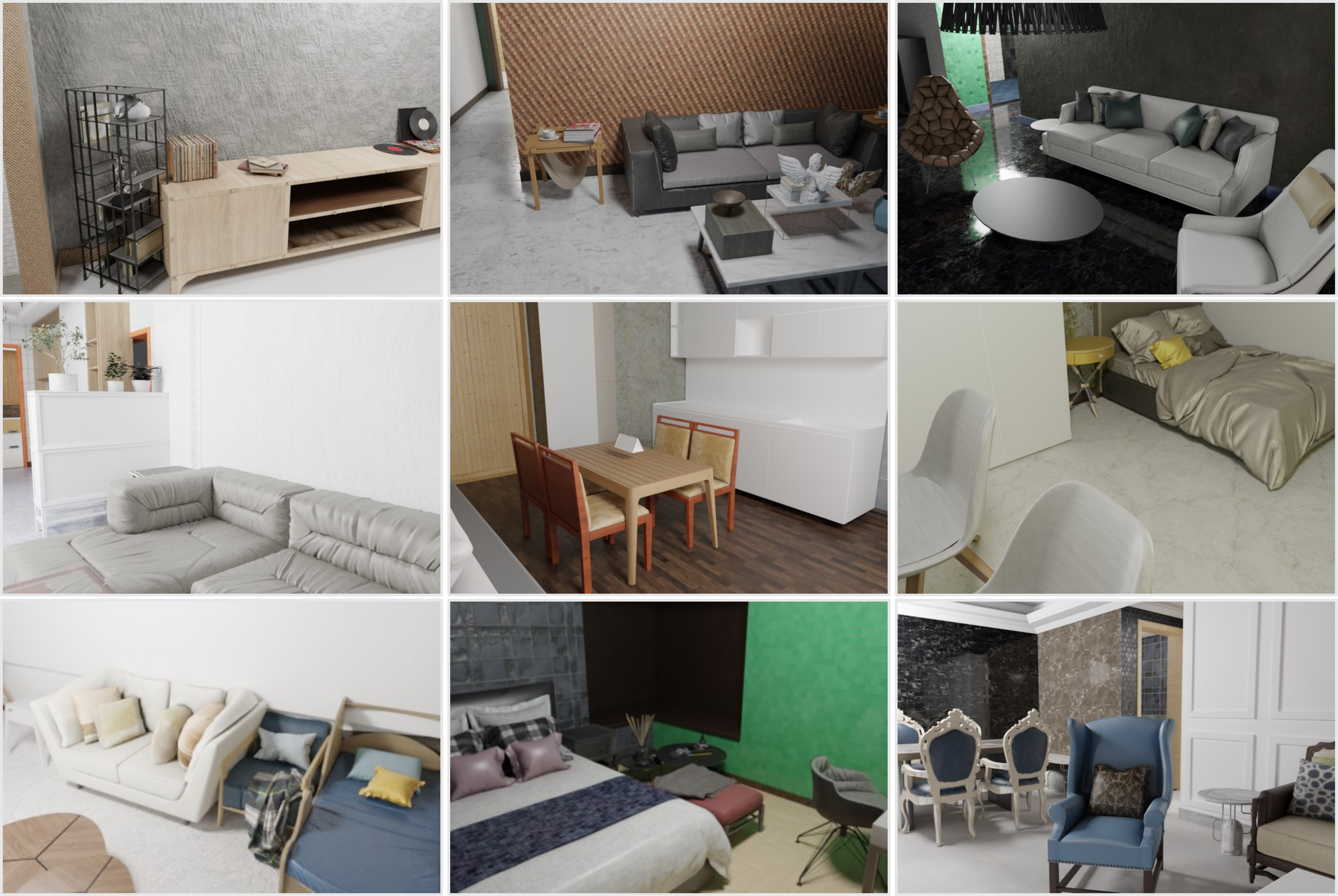}
    \caption{Some indoor scenes rendered in our synthetic dataset.}
    \label{fig:pbrs_img}
\end{figure}

We created a novel synthetic dataset of indoor scenes (see Fig. \ref{fig:pbrs_img}) to train our network, which is based on the 3D-Front~\cite{fu20203dfront} public interior dataset with over 60k distinct house layouts. 
To the best of our knowledge, there is no publicly available dataset containing completely annotated lighting (with factorized shading and reflectance, separated direct/indirect and diffuse/specular components) and geometry features that adapt well to our task. Our dataset will be released publicly online upon publication to fill the absence of a high-quality dataset for learning-based shading prediction tasks.

Our synthetic dataset contains over 30,000 high dynamic range (HDR) images rendered from 1.5k indoor layouts from 3D-Front, each with pixel-wise annotations of lighting, texture, and geometry features. While all textures in 3D-Front are restricted to diffuse materials, and some surfaces have no materials, we also assign physically based materials to the surfaces, which provide highly diverse visual effects and improve visual realism compared to similar synthetic datasets as~\cite{song2017suncg,zhang2017pbrs}.
Note that our dataset also applies to many other computer vision tasks, such as indoor scene understanding, inverse rendering, intrinsic decomposition, etc.  Please see the supplementary document for more details and illustrations about material assignment and dataset generation.

To create photo-realistic rendered images with less manual effort, we use a custom data generation pipeline based on Blender Cycles~\cite{blender}, which can be summarized as the following steps:

\noindent
\textbf{Texture assignment}
To achieve more realism on our synthetic dataset, we use an open-source CC0 texture library~\cite{cc0textures} which includes over 1,300 types of physically based materials covering various surfaces such as wood, metal, glass, etc. We randomly assign certain materials based on the object type in a rule-based style, using a similar strategy with~\cite{denninger2019blenderproc}. For example, for each wall surface, we uniformly sample one material among all wood, marble, or concrete-type materials. This rule-based technique for random texture assignment significantly reduces the effort of manually assigning the textures for each object while preserving sufficient consistency and diversity for indoor layouts.   

\noindent
\textbf{Light sources placement}
We randomly sample the placement of light sources. To approximate direct lighting at real-time performance, we use directional and point/spherical light sources only. We place point and spherical light sources for each room at random positions near the ceiling, and let all spherical light sources' radius and intensity vary within a certain range to acquire more diverse visual appearances.   

\noindent
\textbf{Viewpoint sampling}
We uniformly sample camera poses within the scene, then heuristically filter those that are considered to be of low quality. Specifically, we examine the statistics of depth distribution within each image, particularly thresholding the mean $\mathrm{E}(\mathbf{D})$ and variance $ \mathrm{Var}(\mathbf{D})$ of depth values. Part of images with incorrect statistics (e.g., a low $\mathrm{E}(\mathbf{D})$ or too many dark pixels) will be discarded, which is often the case where cameras are partially/completely occluded by surfaces or nearby objects.

\section{Experiments and Evaluations}
\label{sec:results}
\begin{figure*}[htb]
    \centering
    \includegraphics[width=\linewidth]{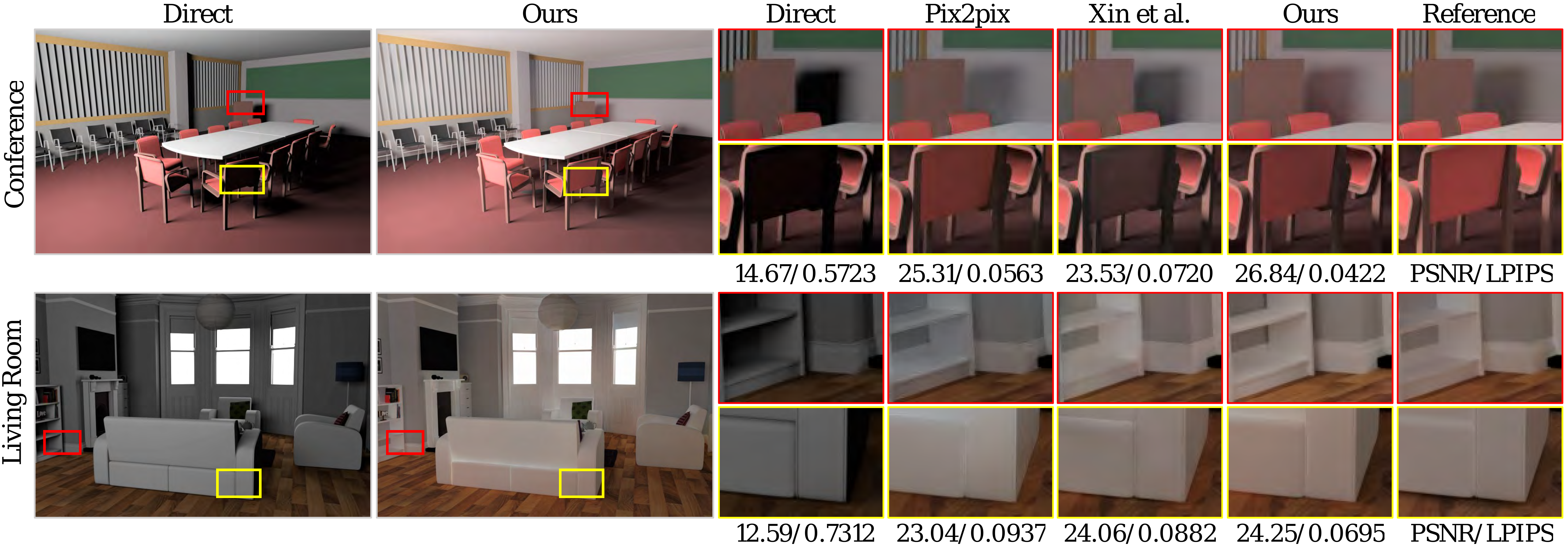}
    \caption{With regard to global illuminations, our approach outperforms other methods in terms of PSNR and LPIPS, and our approach produces results that are visually closer to the reference. }
    \label{fig:compare2}
\end{figure*}

\begin{figure*}[htb]
    \centering
    \includegraphics[width=\linewidth]{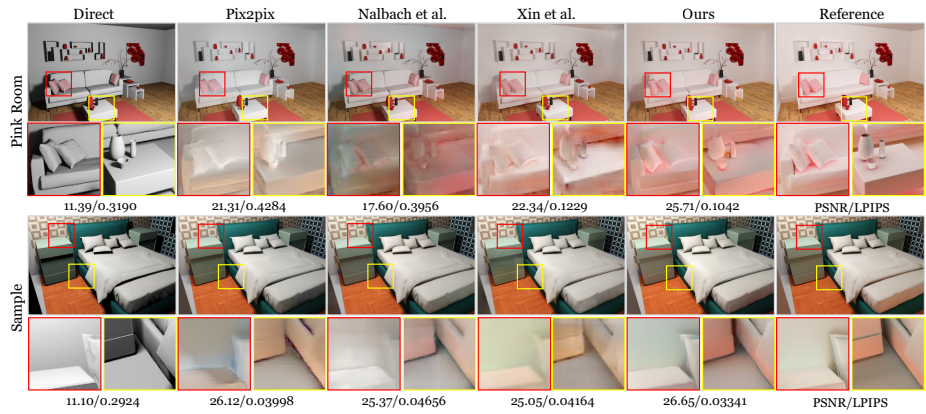}
    \caption{Upper of each scene: full rendering (direct + indirect + texture); lower of each scene: untextured indirect component with zoom-in regions (demodulated from reflectance). our approach outperforms other methods in terms of PSNR and LPIPS.  }
    \vspace{-0.1in}
    \label{fig:compare}
\end{figure*}

\subsection{Implementation Detail}
\label{sssec:implementation}

We implement our network model using PyTorch. For the output of each convolutional layer, we apply Leaky ReLU activation, except for the last layer where an exponential activation is used. For both the generator and discriminator, we use Adam optimizer~\cite{kingma2014adam} for optimization and Xavier~\cite{glorot2010xavier} for network initialization. Dropout~\cite{srivastava2014dropout} is used in up-sampling layers to avoid over-fitting. The model is trained for 80 epochs using over 50 hours. Model performance is measured on an RTX 3090, using $768 \times 512$ (3:2 ratio) as the standard resolution for network inference to synthesize image. Please refer to the supplemental material for our code implementation and the pre-trained model, which will be released online along with the training dataset.

\subsection{Results and Discussions}
\label{sssec:results}

\begin{figure*}[!ht]
    \centering
    \includegraphics[width=\linewidth]{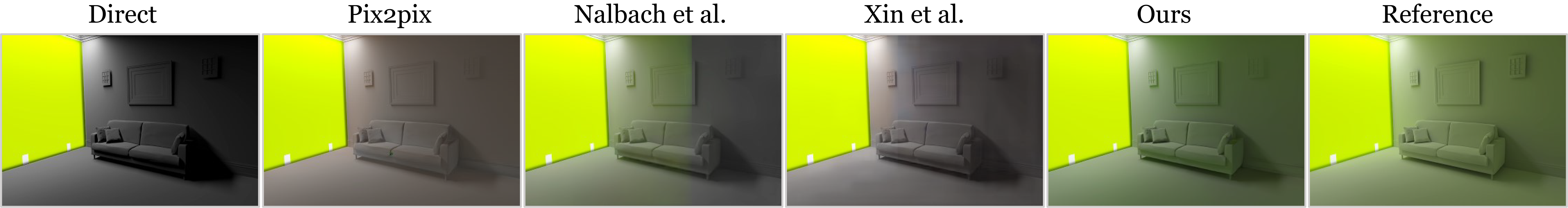}
    \caption{Capability of capturing long-range indirect light of different methods. A green wall on the left is treated as an indirect light source, lit by an area light facing the wall. Our approach synthesizes the better result closer to the reference.
    }
    \label{fig:longrange_ovo}
\end{figure*}

\begin{table}[!bp]
\centering
\begin{center}
\caption{Validation on our test set demonstrates that our method achieves superior performance compared to existing approaches. "Ambient" refers to applying a constant, uniform incident radiance to approximate indirect light.}
\label{tab:compare_perf}
\resizebox{0.65\linewidth}{!}{%
\begin{tabular}{lcccc}
    \toprule 
    \centering
     & \textbf{Params.} & {\bfseries SSIM $\uparrow$ } & \textbf{ PSNR $\uparrow$ } & \textbf{ LPIPS $\downarrow$ } \\ 
    \midrule
    \myrowcolour
    Direct Illumination & - & 0.8676 & 22.17 & 0.0804 \\
    Ambient & - & 0.9455 & 26.15 & 0.0832 \\
    \myrowcolour
    Pix2pix~\cite{isola2017image} & 41.83M & 0.9611 & 28.88 & 0.0352  \\
    Nalbach et al.~\cite{nalbach2017deep} & 7.75M & 0.9564 & 27.95 & 0.0391 \\
    \myrowcolour
	Xin et al.~\cite{xin2021lightweight} & 3.95M & 0.9658 & 30.28 & 0.0312 \\
    \textbf{Ours} & 14.42M & \textbf{0.9743} & \textbf{31.88} & \textbf{0.0203} \\
    \bottomrule 
\end{tabular}
}
\end{center}
\end{table}

We use three evaluation metrics to compare the performance of different methods, including SSIM, PSNR, and LPIPS~\cite{zhang2018lpips}. The LPIPS metric calculates perceptual similarity based on high-level features in a pre-trained AlexNet~\cite{krizhevsky2012alexnet}, which is more robust to noise and better resembles human choice. We compare two recent works on global illumination prediction, including Nalbach et al.~\cite{nalbach2017deep} and Xin et al.~\cite{xin2021lightweight}. 
Our method adopts a neural rendering approach based on shading space, predicting indirect illumination from screen space to achieve real-time HDR global illumination estimation. Its core idea is to efficiently simulate long-range indirect light transport through an image-to-image learning paradigm, without relying on explicit geometric modeling or physical path integration, thus belonging to a fast, screen-space neural rendering framework. In contrast, \cite{zheng2024neural} focuses on 3D explicit scene modeling and object-level light transport representation. It implicitly models the illumination contribution of each object in the scene, superposing them in latent space to solve for global illumination. These methods differ fundamentally in technical routes, input spaces, and problem formulations. Therefore, such scene-space neural rendering methods \cite{zheng2024neural,su2024dnr,su2025vfnr,hadadan2021neural} are not included for direct comparisons.

Note that Nalbach et al.~\cite{nalbach2017deep} directly reconstructs the global illumination image, while Xin et al.~\cite{xin2021lightweight} predicts the indirect component, similar to our task setup. For Nalbach et al.'s implementation, we use the "mono" version they described.
We train all models on our dataset since the dataset they used is not publicly available. We also compare our method with the commonly used models in image-to-image translation or image style transfer tasks, including Pix2pix~\cite{isola2017image} and ResNet~\cite{he2016deep} with the same training setup and loss function as our model. All evaluation results are obtained on our test set, including 1000 rendered frames from our synthetic dataset and about 200 frames from public scenes we collected, e.g., \textbf{PinkRoom}, \textbf{LivingRoom}, \textbf{Gallery} and \textbf{Conference}. The averaged results of the evaluation metrics are shown in Tab. \ref{tab:compare_perf}.

As can be seen, our model consistently outperforms all previous works and different network models on the three metrics. While the SSIM metric has a somewhat smaller difference among different methods, there is a huge gap between our method and other methods on the perceptual metric LPIPS, implying more natural illumination produced by our method. 
Visualizations of the whole predicted illumination are shown in Fig. \ref{fig:compare2}. We further look into the predicted indirect component (demodulated from texture) in Fig. \ref{fig:compare}, which is more intuitive for comparing the performance of different methods.
As observed, Xin et al.'s method used a lightweight model and tended to produce over-smoothed illumination on high-frequency regions, while the deeper models (Pix2pix and ResNet) using more parameters sometimes produced noisy and distorted results. 
Our method strikes a balance between model complexity and image quality, producing more plausible results at a reasonable time budget ($12\text{ms}$ at $768\times 512$), compared to the methods of Nalbach et al.~\cite{nalbach2017deep} ($6\mathrm{ms}$) and Xin et al.~\cite{xin2021lightweight} ($13\mathrm{ms}$).

Another observation is that our model can accurately handle the color of indirectly reflected light, while all other methods cannot in most cases. In our experiments, smaller networks (e.g., Xin et al.'s method) tend to produce over-smoothed results. Whereas, the deeper generative models (e.g., Pix2pix and ResNet), however, are easy to overfit and produce the averaged color (i.e., the most frequently occurred one) within training data, resulting in erroneously predicted colors. This problem remains if other color models are used, e.g., CIE XYZ or HSV. We further verify this by selecting frames that feature interreflections between differently textured surfaces (see Fig. \ref{fig:showcolor}). We attribute this success to the design of the monochromatic generator, which will be further discussed in the next section.

\subsection{Validation}
\label{sssec:validate}

To verify the effectiveness of the design choices in our model, we conduct ablation studies by modifying different components of our model and comparing their performance with the full model.

\begin{figure}[!tb]
    \centering
    \includegraphics[width=0.7\linewidth]{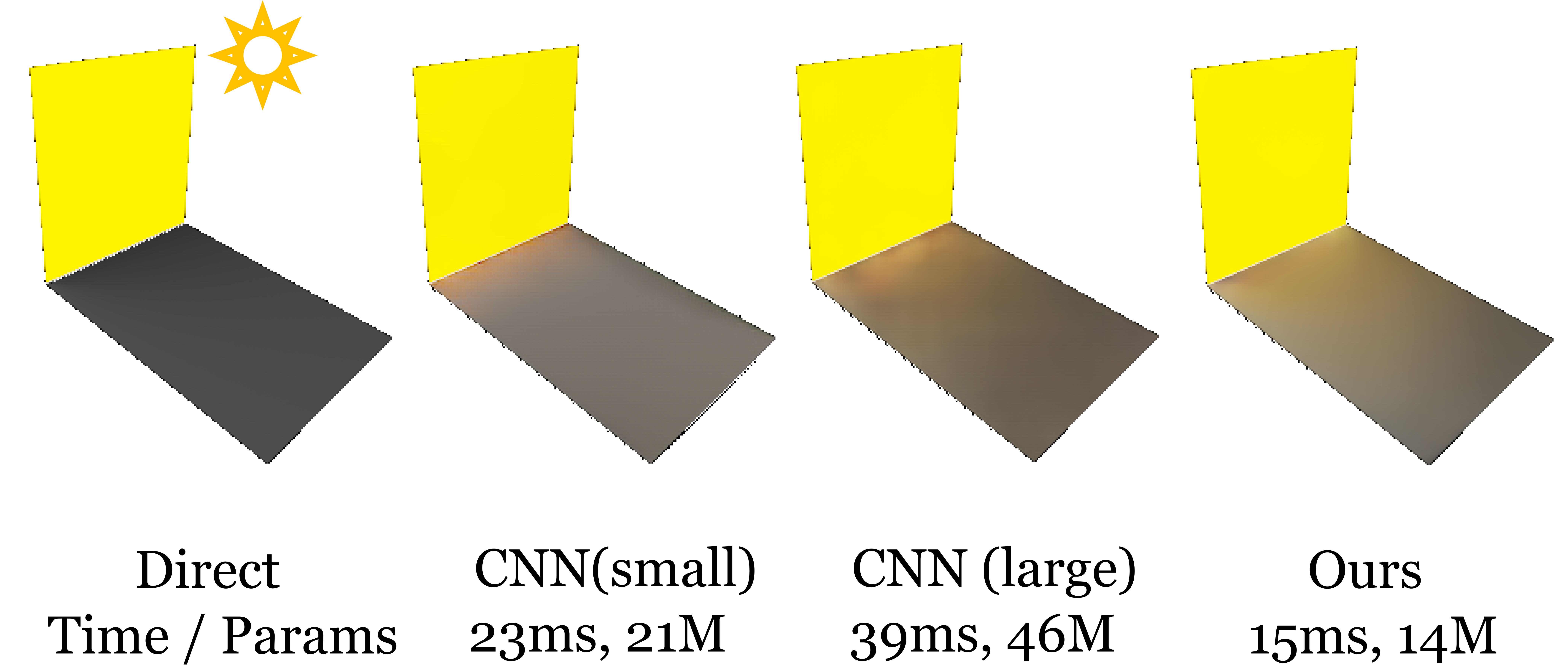}
    \caption{Our GFA module effectively captures distant indirect light compared to stacks of convolution operations. The object covers an area of about $512 \times 512$ pixels, larger than the receptive field of the \emph{CNN(small)} model. }
    \vspace{-0.1in}
    \label{fig:longrange}
\end{figure}
\begin{figure}[!htb]
    \centering
    \includegraphics[width=0.7\linewidth]{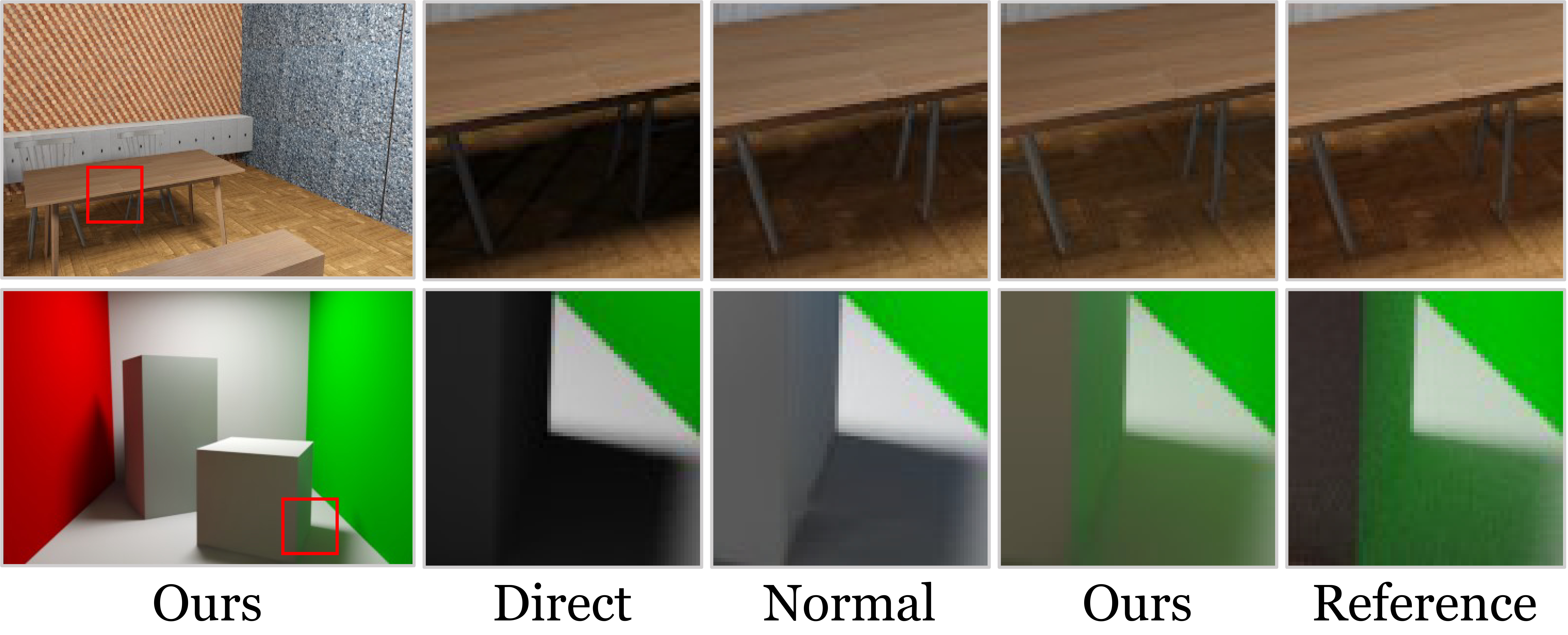}
    \caption{Validation of our monochromatic design. Our approach handles the colored indirect illumination better, especially in the shadowed regions. "Normal" refers to the generally used network with RGB inputs. }
    \vspace{-0.1in}
    \label{fig:monoab}
\end{figure}
\begin{figure*}[!ht]
    \centering
    \includegraphics[width=\linewidth]{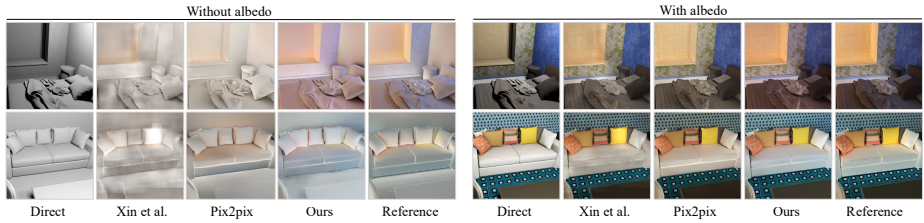}
    \caption{Diffuse interreflections. Both the untextured (left) and textured (right) indirect components are shown. Our model can accurately capture the colored indirect light reflected from nearby surfaces, while the other methods cannot. }
    \label{fig:showcolor}
\end{figure*}
\noindent
\textbf{Effectiveness of global geometry-aware feature aggregation} As discussed in Sec. \ref{sec:method}, we exploit the modified multi-head attention mechanism for guiding global feature aggregation based on geometry features. This allows modeling long-range dependencies for any two locations in the image space, which is inherently impractical for fully convolutional networks (FCN) that prior works used. We validate this with a simple scene containing one major indirect light source (see Fig.~\ref{fig:longrange_ovo}), whose horizontal span exceeds the receptive field of the compared models. We show that previous works have difficulty capturing distant indirect light sources due to the locality of convolution operations, whereas our method efficiently models long-range dependencies with a feature aggregation module. Please refer to our supplemental video for more comparisons under dynamic lighting.

We also validate the efficiency of our method by simply stacking more convolutional layers using a similar setup (see Fig.~\ref{fig:longrange}). We compare our method to the variants that use convolutional layers to replace the feature aggregation module within the network bottleneck (i.e., two models with 7/9 layers in both their encoders and decoders). All variants use the monochromatic design and the same training strategy for a fair comparison. We show that the 7-layer model is unable to capture distant indirect lighting. While it is possible to enlarge the receptive field by using a deeper stack of convolutional operations (e.g., the 9-layer model), it brings extra computational overhead and optimization difficulty while still performing worse than our proposed feature aggregation method. We also note that the prior works generally perform worse than the 7-layer model in capturing long-range dependencies since they all have smaller receptive fields. 

We then investigate the effectiveness of the multi-head extension by changing the number of attention heads (see the middle section of Tab. \ref{tab:monoab}). We show that the model performs better by increasing the number of heads, while introducing little computational overhead or extra parameters (0.024 million parameters per head). We thus believe it important to attend to different subspaces of the features in bottleneck layers, where the features have a down-sampled spatial resolution but with high dimensionality.

\noindent
\textbf{Effectiveness of monochromatic design}
We regard the monochromatic design as the main reason that helps the model capture the correct color from indirectly reflected light (see Fig. \ref{fig:showcolor}). To further validate this, we implement an alternate version of our model, taking in the RGB inputs and directly reconstructing the RGB output, namely the \textbf{Normal} network. Both versions of the network use the same architecture except for the image reconstruction process. As shown in Fig. \ref{fig:monoab}, our model is significantly more capable of capturing the colored indirect lighting. This is most visible in the regions consisting of indirect illumination only (e.g., the shadowed areas in CornellBox). Note that while Nalbach et al.~\cite{nalbach2017deep} also suggested the independent inference for each channel, they simply take it in a completely independent manner, resulting in visual artifacts like color-shifting (see Fig.~\ref{fig:longrange_ovo} in the main text and also Fig. 5 in Nalbach et al.~\cite{nalbach2017deep}) due to their network's poor ability to preserve linearity (i.e., $\mathbf{Net}(\alpha x) \ne \alpha\mathbf{Net}(x)$). In contrast, our network design takes several strategies to avoid this downside, including reusing shared geometry features, applying an inter-channel perceptual loss, and an effective data augmentation strategy for HDR training (see Sec.~\ref{sec:method}). 

Moreover, to justify the potential performance loss of the monochromatic design (introduced by the parallel inference of three channels, see Sec. \ref{sec:method}), we implement a mini version of the monochromatic network $\textbf{Ours (mini)}$, using the same network architecture as $\textbf{Ours}$'s but reducing the number of channels in intermediate layers to roughly half the model size. As shown in the first section of Tab. \ref{tab:monoab}, the $\textbf{Ours (mini)}$ still performs better than \textbf{Normal} with similar computational cost and a much smaller amount of parameters. As discussed in Sec. \ref{sec:intro}, we believe the success of the monochromatic network may result from three major reasons: 1) eliminating the redundant connections (correlations) between different color channels; 2) reducing the high-dimensional space of RGB input/output by introducing the prior that the rendering process of each color channel (or wavelength) is independent; thus 3) making the network more robust to the inequality and bias in the color distribution of training data.
To further verify the superiority of our monochromatic model, we introduce scenes illuminated with colored lights that are not present in our training data (see Fig. \ref{fig:teaser} and Fig. \ref{fig:colorlight}). Our monochromatic model can still produce plausible outputs without further training or fine-tuning, whereas the normal models generally cannot achieve this with no (or only a few) training data in this domain.

\begin{table}
\centering
\begin{center}
\caption{We use ablation experiments to show the effectiveness of the multi-head attention mechanism (via changing the number of attention heads) and the monochromatic design, respectively. }
\label{tab:monoab}
\resizebox{0.7\linewidth}{!}{%
\begin{tabular}{lccccc}
    \toprule 
     & \textbf{Params.}  & \textbf{Time(ms)}  & \textbf{ SSIM $\uparrow$ } & \textbf{ PSNR $\uparrow$ } & \textbf{ LPIPS $\downarrow$ } \\
    \midrule 
    \myrowcolour
    $\textbf{Normal}$ & 15.17M & \textbf{9.15} & 0.9625 & 31.49 & 0.0316  \\
    $\textbf{Ours(mini)}$ & \textbf{8.76M} & 9.76 & 0.9663 & 31.62 & 0.0271 \\
    \midrule
    \myrowcolour
    $\textbf{Ours-1Head}$ & 14.29M & 11.55 & 0.9545 & 31.13 & 0.0247 \\
    $\textbf{Ours-4Heads}$ & 14.35M & 11.73 & 0.9419 & 30.92 & 0.0237 \\
    \midrule
    \myrowcolour
    $\textbf{Ours (No $\mathcal {L}_p$ )}$ & 14.42M & 12.17 & 0.9705 & 31.52 & 0.0232  \\
    $\textbf{Ours (No $\mathcal {L}_a$ )}$ & 14.42M & 12.17 & 0.9688 & 31.42 & 0.0240 \\
    \midrule
    \myrowcolour
    $\textbf{Ours}$  & 14.42M & 12.17 & \textbf{0.9743} & \textbf{31.88} & \textbf{0.0203} \\
    \bottomrule 
\end{tabular}
}
\end{center}
\end{table}

\begin{figure}[!htb]
    \centering
    \includegraphics[width=0.7\linewidth]{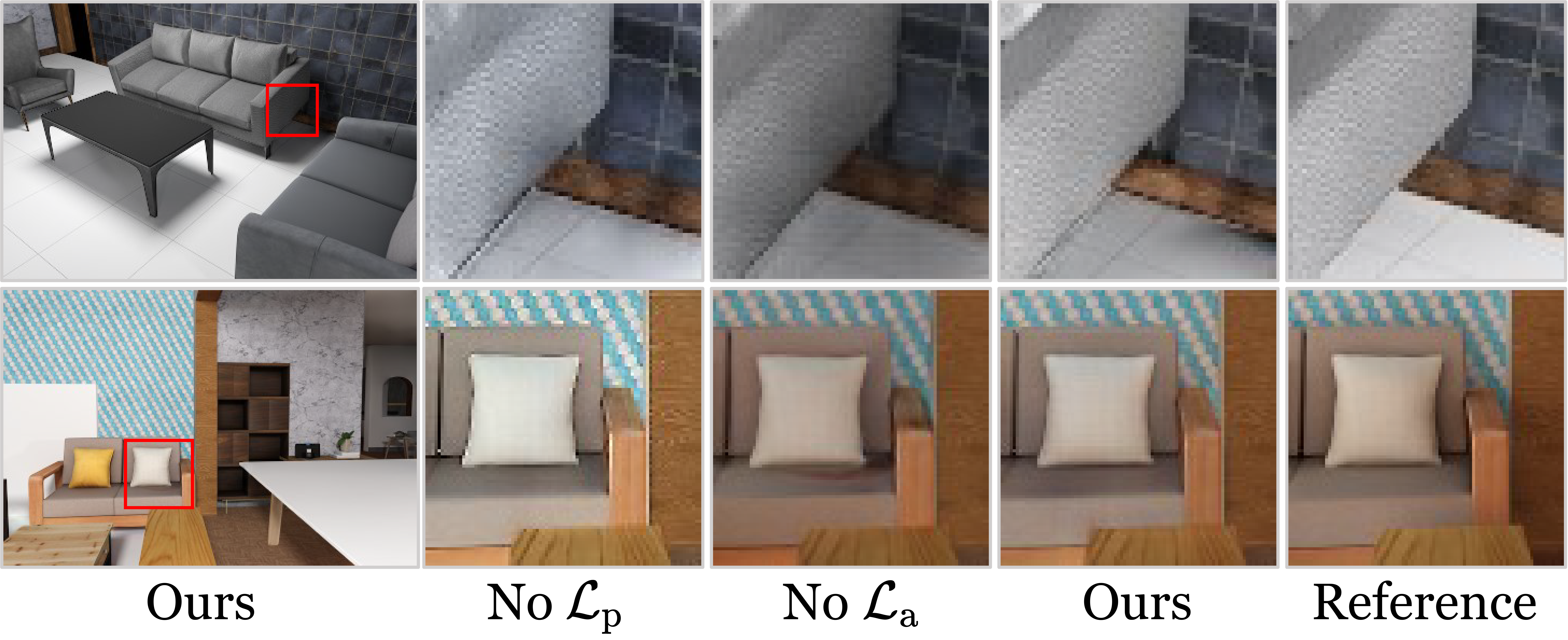}
    \caption{Using different loss functions (indirect component). The predicted color tends to be erroneous without perceptual loss (No $\mathcal{L}_{\text{p}}$), and the results are blurry without adversarial training (No $\mathcal{L}_{\text{a}}$). Both variants tend to produce more visual artifacts. }
    \vspace{-0.1in}
    \label{fig:lossab}
\end{figure}

\begin{figure}[!htb]
    \centering
    \includegraphics[width=0.65\linewidth]{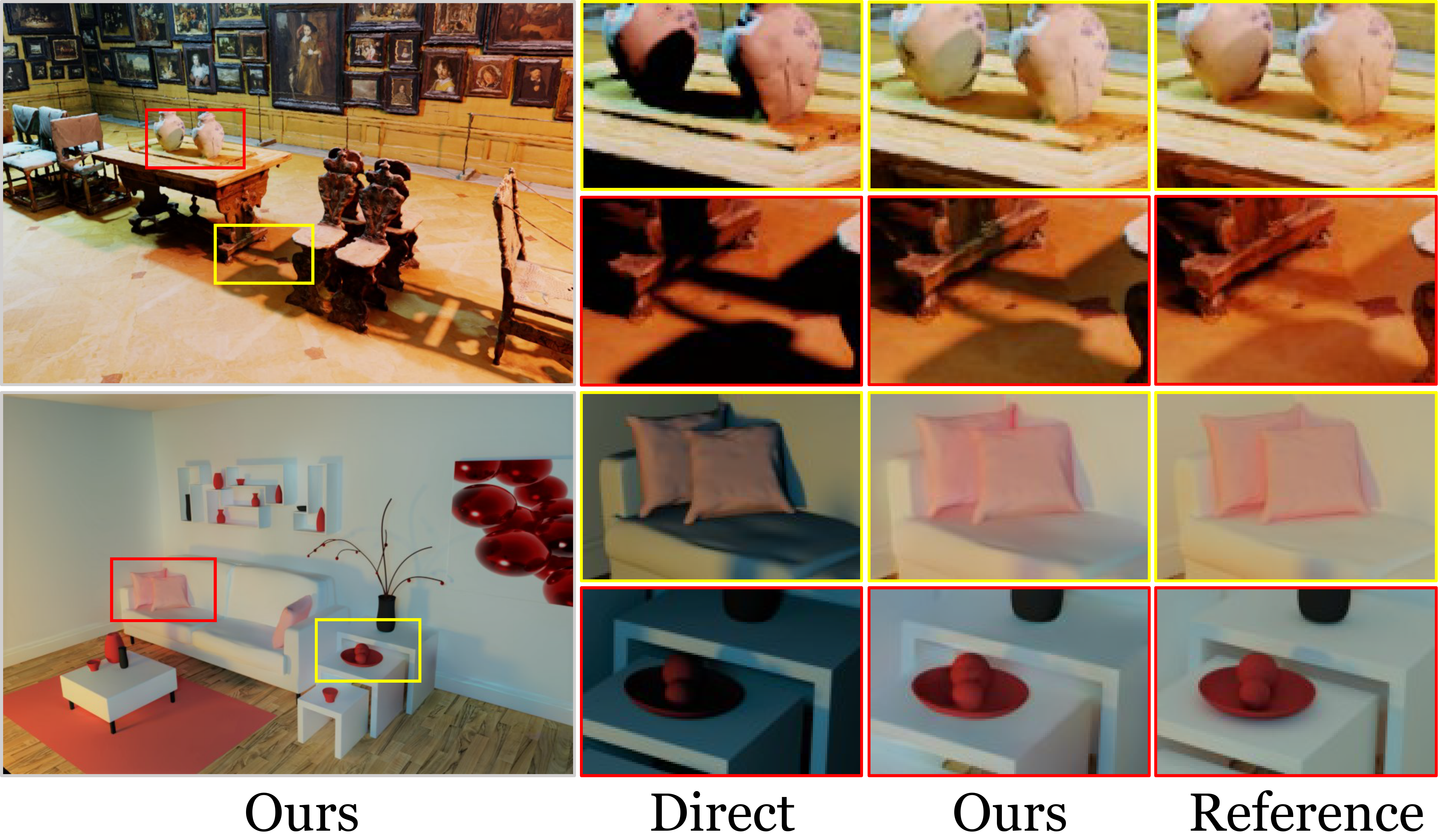}
    \caption{Scenes lit by the colored light source that are not in the training data. Our approach is well generalized to colored lights due to the independent prior of each color channel.}
    \vspace{-0.1in}
    \label{fig:colorlight}
\end{figure}

\noindent
\textbf{Effectiveness of perceptual and adversarial loss}
We validate the effectiveness of the perceptual and adversarial components in our loss function by removing each of them and comparing their performance with the full model. As shown in Fig. \ref{fig:lossab}, the model produces color-shift artifacts without the perceptual term, whereas the outputs are blurry without the adversarial term. We thus assert that the adversarial loss helps reconstruct the high-frequency details, and the perceptual loss improves the visual consistency across channels.
Moreover, we found that the perceptual metric (e.g., LPIPS) becomes much worse when removing either of the loss terms, since both the adversarial discriminator and the pre-trained VGG can capture the structurally correlated image patterns at a higher abstract level. In contrast, classic metrics like L1 and L2 simply assume no interconnections between pixels, which are usually vulnerable to noise and local artifacts. We thus use a mixture of pixel-wise loss and perceptual loss, along with the adversarial training strategy, to guide our model to produce more plausible indirect illumination.

\subsection{Discussion}
\label{sssec:discuss}
\noindent
\textbf{Network generalization}
As discussed in Sec. \ref{sssec:results}, our model can be generalized to the publicly collected scenes, as well as the colored light sources, which are not present in our training data. Based on this, we further show that our model can also generalize to dynamic lighting with more complex light sources, such as area lights and image-based environment lighting (acquired using Monte Carlo integration since they cannot be effectively approximated), with no obvious downgrade of performance. We regard this success of generalization as a result of multiple reasons, including 1) the use of the monochromatic design makes the network unaffected by color shifting and color variances; 2) the training setup of learning shading instead of scattered radiance makes the learning process easier; and 3) the diversity of our synthetic training data in terms of materials, geometry, and viewpoints. Please see our supplemental video for network performance under continuous frames and dynamic lighting. 

\noindent
\textbf{GPU-accelerated ray tracing alternatives}
While there are multiple existing methodologies for efficient specular or glossy reflections in different use cases \cite{mcguire2014efficient,hirvonen2019accurate}, the diffuse indirect illumination needs more sophisticated processing due to its uniform scattering nature (resulting by its hemispherical diffuse BRDF lobe, instead of glossy or Dirac delta ones). Recently, the development of GPU ray tracing acceleration offered new capability for real-time indirect illumination, among which sparsely traced results with denoising is a promising technique. 

\begin{figure}[!tbp]
    \centering
    \includegraphics[width=0.7\linewidth]{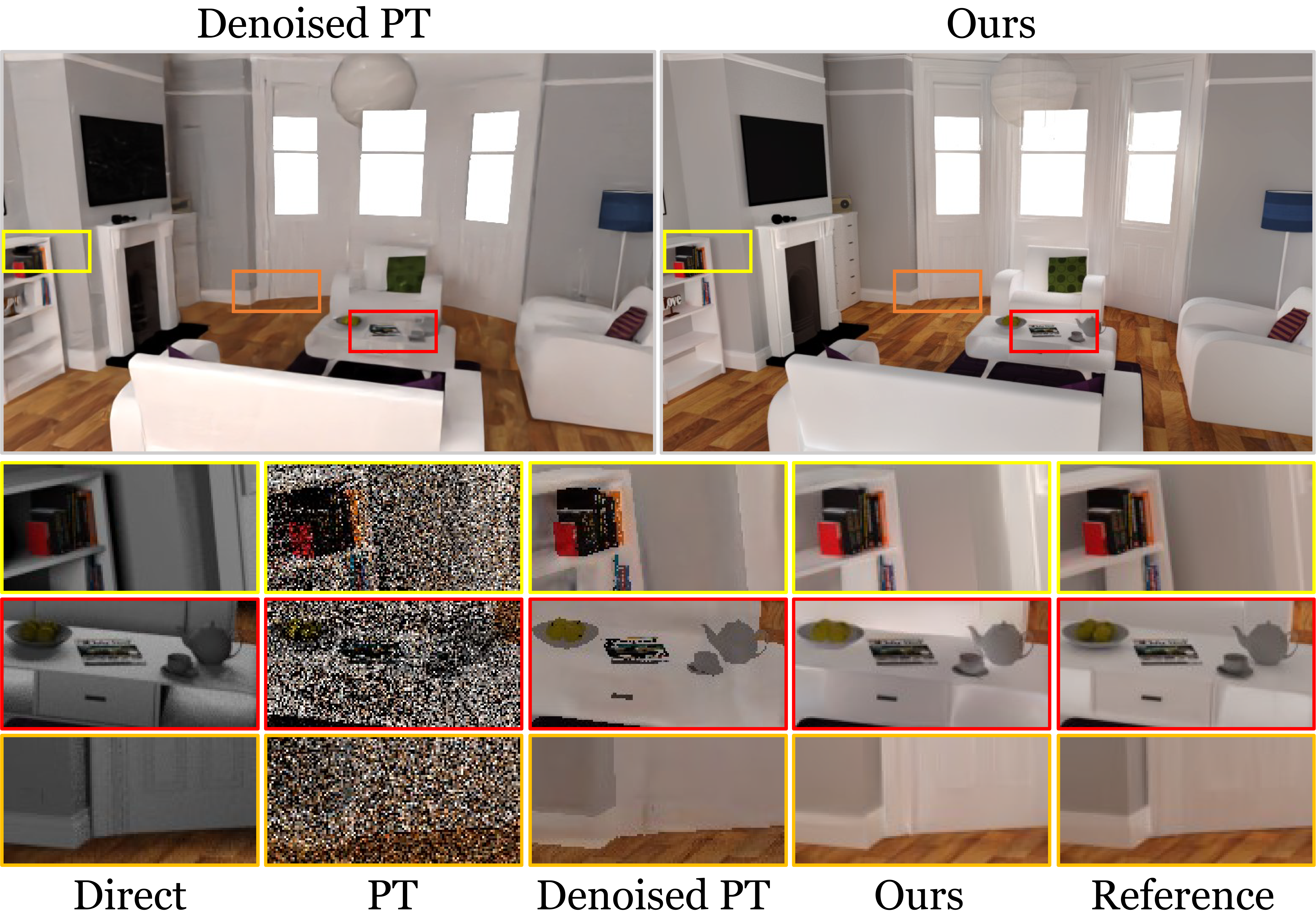}
    \caption{Our method vs. denoised path-tracer (PT). Path tracing ($1\mathrm{spp}$) with next event estimation (NEE) is used, which is then denoised with the AI-Accelerated denoiser in OptiX\textsuperscript{\texttrademark}~\cite{chaitanya2017mc}. Our method is closer to the reference in many scenarios in terms of the predicted indirect component while using less time consumption than the Denoised PT.}
    \vspace{-0.1in}
    \label{fig:denoisept}
\end{figure}

Thereby, we compare our method with the GPU-accelerated ray tracing+denoising alternative, which often uses larger time consumption than our method. As shown in Fig.~\ref{fig:denoisept}, we found that the denoised results suffer from loss of high-frequency details, implying that our method might still be beneficial for modern rendering applications. Nonetheless, we think that the ray tracing techniques could be used to compensate for the absence of the out-of-screen features in our method, which might be an interesting avenue for future work.

\noindent
\textbf{Rendering Pipeline Integration}
Our model can be integrated into an interactive rendering pipeline, given that the direct illumination is relatively simple and can be approximated with real-time algorithms~\cite{fernando2005pcss,heitz2016real}. 
By doing this, all the inputs are generated on-the-fly within GPU memory, which can be immediately sent to the neural networks without the additional cost of memory transfer operations. However, the practicality of our method is limited in the following aspects, which could be improved straightforwardly:

\begin{itemize}

    \item \emph{Performance}. Since the image's resolution will heavily affect the time cost for inference, our model works at $768 \times 512$ and can generate outputs in $12 \mathrm{ms}$. For those VR applications with a higher resolution requirement (e.g., monocular at 1800$\times$1920 on Meta Quest), We can accelerate the inference by bilateral up-sampling techniques like~\cite{kopf2007joint} at an even lower resolution, which could upscale the predicted indirect illumination at a low time budget since the indirect component has few high-frequency details. 
    
    \item \emph{Temporal coherence}. Our method may produce perceivable flickering under drastic viewpoint or geometry changes, which is common when applying learning-based methods to video. This can be alleviated with temporal anti-aliasing (TAA) techniques.

\end{itemize}

\section{Conclusion, Limitation, and Future Works}
\label{sec:conclusion}

\noindent
We present a screen-space global illumination method based on a novel network design for indirect illumination prediction for any newly-constructed scenarios with complex lighting conditions. Our learning-based method can synthesize realistic images using a time budget close to the real-time rendering of local illumination. Therefore, it paves a feasible path to highly-efficient realistic rendering and can be affordable to many VR/AR applications for realism enhancement. 

While our method could predict plausible indirect illumination, it has several limitations. First, our method has similar limitations to other screen-space methods due to the lack of layered depth features. Second, our method cannot handle those highly glossy or specular surfaces well, which other orthogonal solutions might compensate for in future work. This jointly implies limited practicality for current real-time rendering pipelines, which additional strategies may improve, as discussed in Sec.~\ref{sssec:discuss}.

In the future, we will investigate more effective techniques for indirect illumination prediction while maintaining a relatively lightweight and efficient network architecture for VR applications, especially binocular rendering. Meanwhile, better exploiting 3D spatial features or implicit scene representation (e.g., via multi-view images or temporal coherence) may also be helpful for illumination prediction. 
Our proposed approaches are also potentially beneficial for other tasks. Specifically, the geometry-aware aggregation module might be preferred when geometry-supervised long-range modeling is needed. Moreover, while the monochromatic design is hardly seen in computer vision tasks, it would be interesting to see if it could be applied to various neural rendering tasks, e.g., learning-based Monte Carlo denoising~\cite{chaitanya2017mc} or neural radiance fields~\cite{mildenhall2020nerf}.


\bibliographystyle{splncs04}

\bibliography{template}



\end{document}